\documentclass{appolb}
\usepackage{graphicx}
\usepackage{hyperref}
\usepackage{lineno}

\begin{document}
\title{New constraints on PDFs with CMS data
\thanks{Presented at ``Diffraction and Low-$x$ 2022'', Corigliano Calabro (Italy), September 24-30, 2022.}%
}
\author{L. Alcerro
\thanks{On behalf of the CMS Collaboration.}
\thanks{Supported by the Nuclear Physics program\href{https://pamspublic.science.energy.gov/WebPAMSExternal/Interface/Common/ViewPublicAbstract.aspx?rv=00d4fe0f-48a0-4d4a-baf1-c70867d9e499&rtc=24&PRoleId=10}{ DE-FG02-96ER40981} of the U.S. Department of Energy.}
\address{The University of Kansas}
{ 
\address{}
}
\address{}
}
\maketitle
\begin{abstract}
Recent results on inclusive jet production and production of a W boson in association with a charm quark by the CMS Collaboration are presented in this proceeding. The impact of these measurements on proton PDFs are also discussed. 
\end{abstract}
  
\section{Introduction}
The collinear factorization theorem allows us to separate long- and short-distance contributions by the introduction of Parton Distribution Functions (PDFs) which gives information about the intrinsic hadronic structure hence, they are process independent. Current hadron colliders strongly rely on PDFs, however they represent one of the main sources of uncertainty. 

PDFs are usually extracted from experimental data (Deep Inelastic scattering, Drell-Yan, jets, top quark etc.) and nowadays there are several state-of-the-art PDF sets which use different combinations of available data in their QCD analyses. The CMS experiment has recorded large samples of proton-proton collisions containing different processes directly sensitive to the proton PDFs. 

\section{Inclusive jet production}
\begin{figure}
\begin{center}
\includegraphics[width=\textwidth]{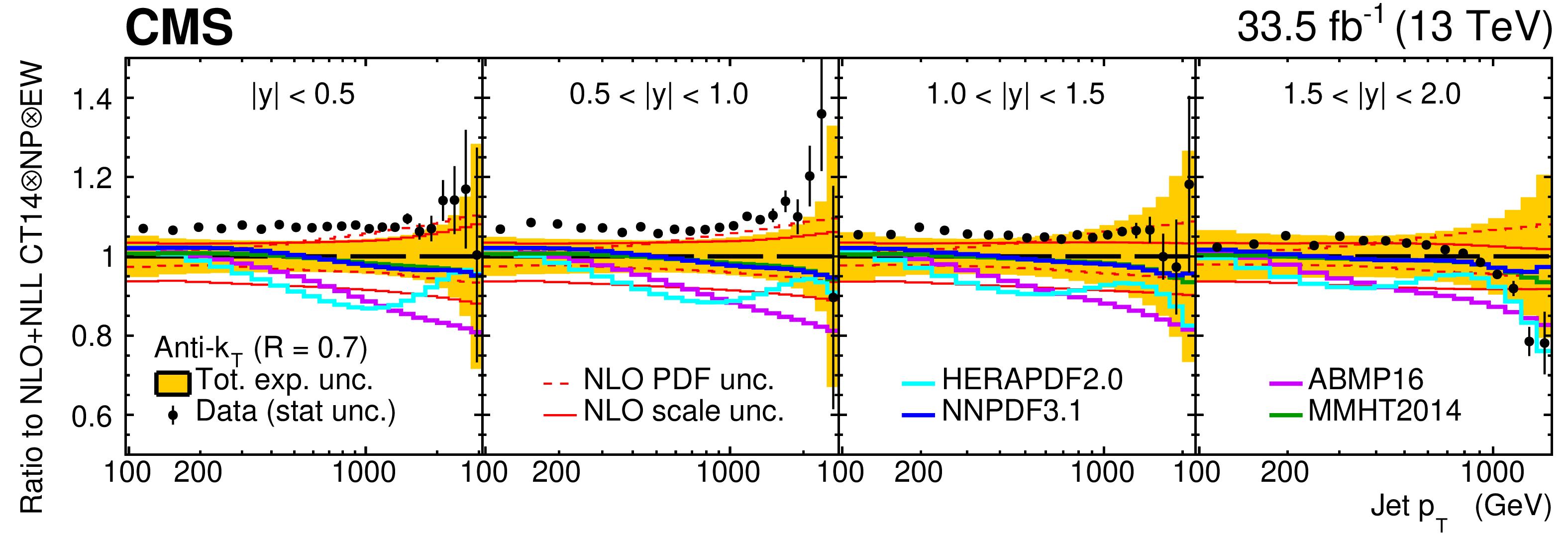}
\caption{Double-differential jet cross sections as a function of jet $p_\mathrm{T}$ and $|y|$ for jets clustered with $R=0.7$.  Plots taken from Ref. \cite{CMS:2021yzl}.}
\label{ratio}
\end{center}
\end{figure}

\begin{figure}
\begin{center}
\includegraphics[width=0.3\textwidth]{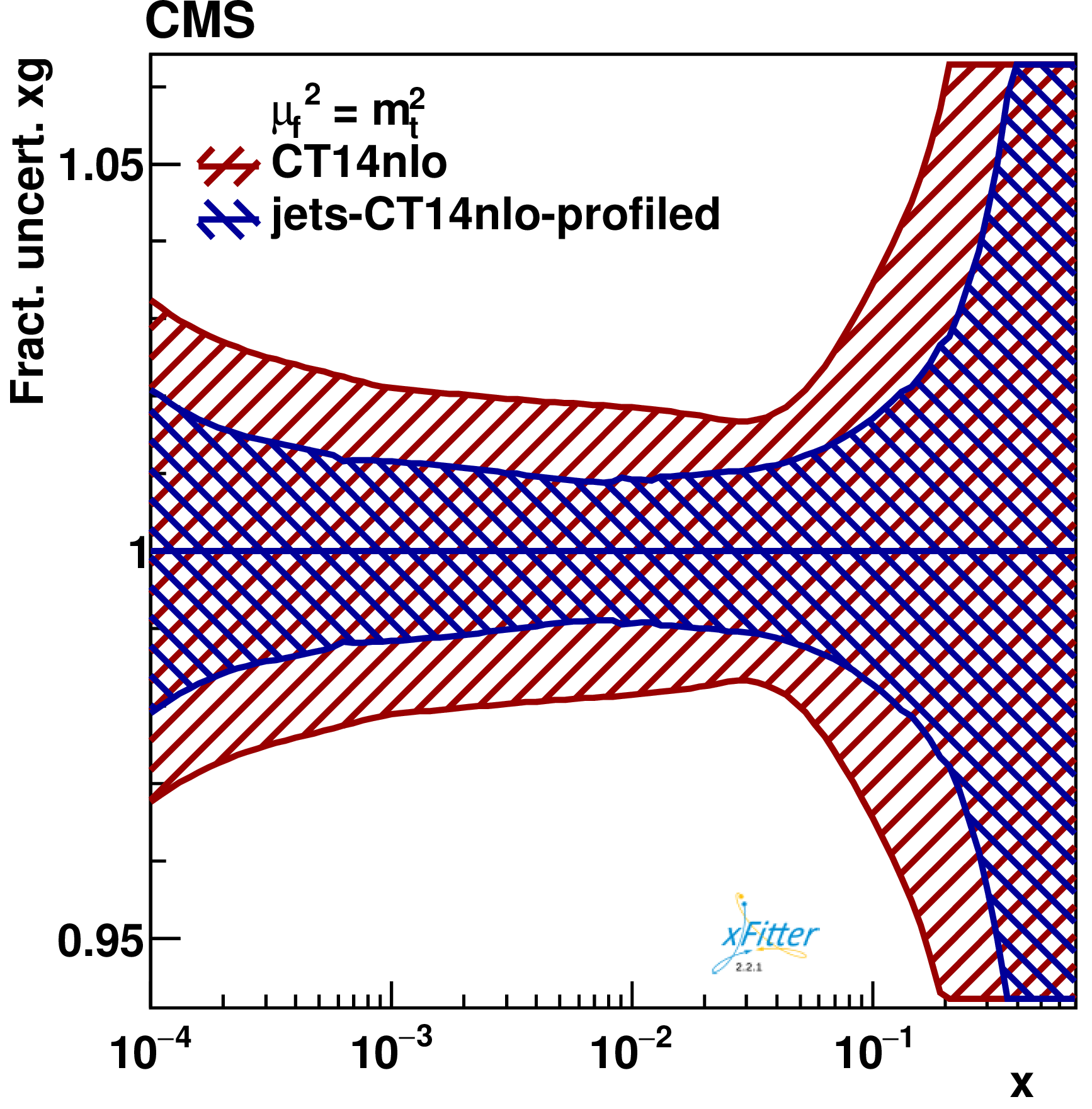}
\includegraphics[width=0.3\textwidth]{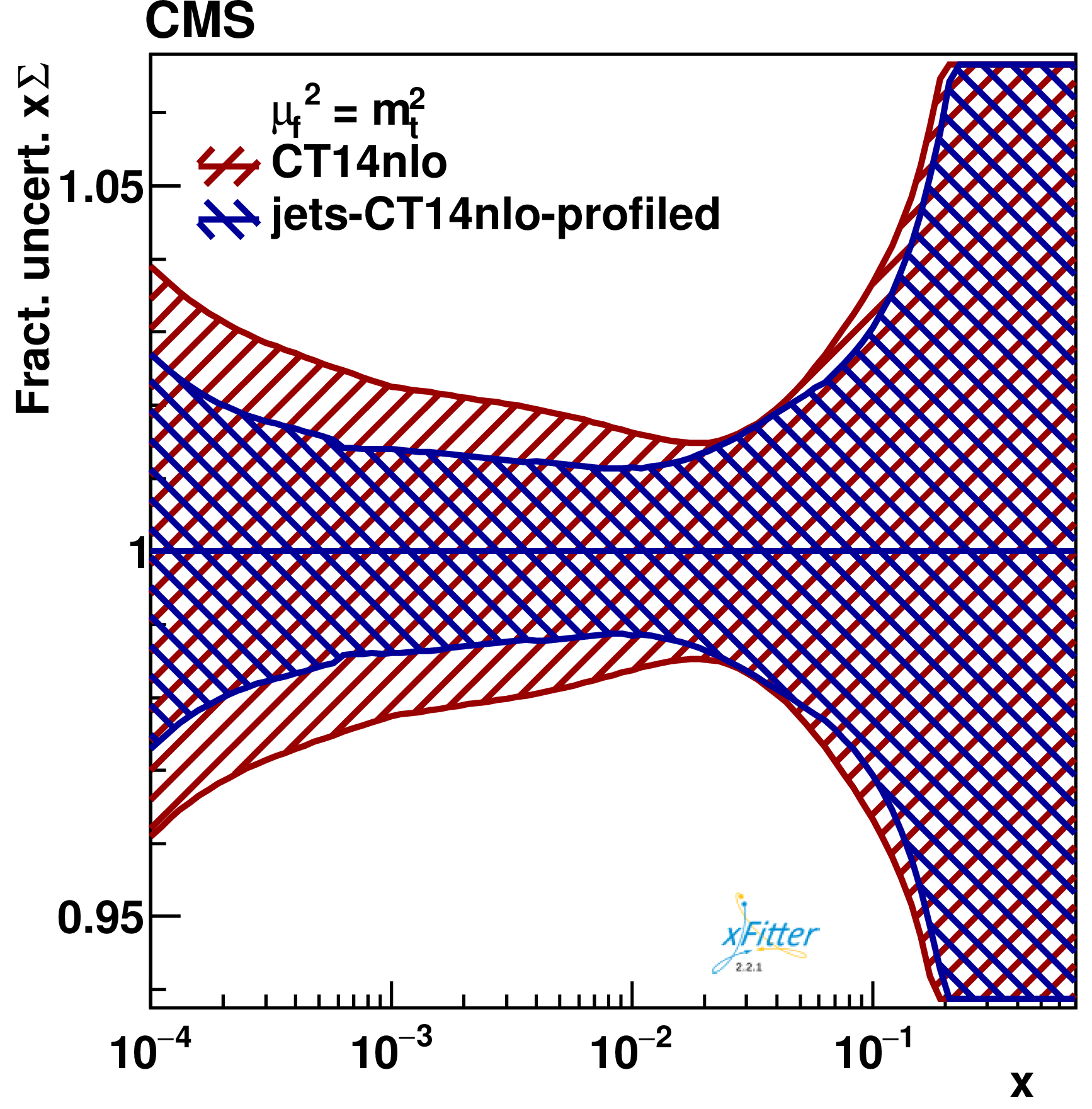}
\includegraphics[width=0.3\textwidth]{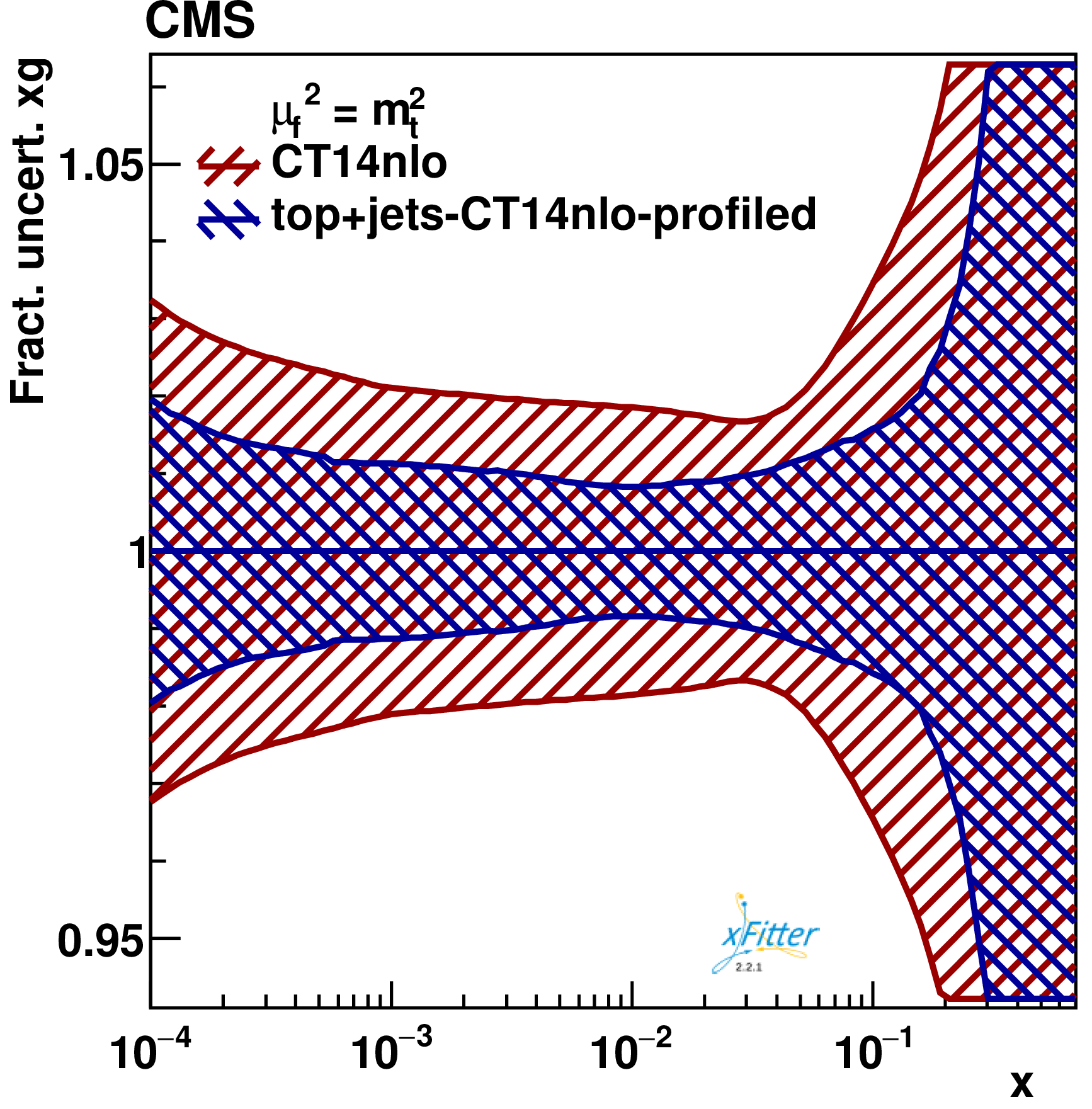}
\caption{Fractional uncertainties in the gluon (left) and sea quark (middle) distributions as a function of $x$. The profiling is performed using CT14nlo PDF at NLO, by using the CMS inclusive
jet cross section at $\sqrt{s}=13$ TeV. The profiling analysis is repeated at NLO including triple-defferential CMS $\mathrm{t\bar{t}}$ cross section (right). Caption Plots taken from Ref. \cite{CMS:2021yzl}.}
\label{profiling1}
\end{center}
\end{figure}

Inclusive jet production has been extensively studied at CMS \cite{CMS:2008xjf} and ATLAS \cite{ATLAS:2008xda} experiments. It provides direct access to the gluon distribution of the proton.
In a recent result \cite{CMS:2021yzl}, the CMS Collaboration has reported a measurement of the inclusive jet production cross section in proton-proton collisions at $\sqrt{s} = 13$ TeV. The data sample corresponds to 36.3 $\mathrm{fb}^{-1}$ (33.5 $\mathrm{fb}^{-1}$) recorded in 2016 for events with jets clustered with the anti-$k_{\mathrm{T}}$ algorithm (AK) with radius $R = 0.4$ $(0.7)$. The inclusive jet double-differential cross section is calculated and shown as ratio to
NLO+NLL prediction with the CT14 PDF set, as in Fig. \ref{ratio}. NLO+NLL predictions obtained with alternative PDF sets are also displayed in Fig. \ref{ratio}. It is worth to notice that neither ABMP 16 nor HERAPDF2.0 include jet measurements and although predictions at low $p_{\mathrm{T}}$ are similar, significant differences are observed at high $p_{\mathrm{T}}$ with these PDFs. These differences come from differences in the gluon distributions of the proton at large value of $x$. 
The impact of this measurement to the proton PDFs and the strong coupling constant is investigated in a QCD analysis using the above mentioned jet cross section for $R=0.7$ together with the DIS cross sections of HERA \cite{H1:2015ubc}. In addition, the normalized triple-differential $\sigma_\mathrm{t\bar{t}}$ cross section \cite{CMS:2019esx} from CMS is used. 
In a profiling analysis at NLO and NNLO, the impact of this measurement is performed using the CT14 PDF sets derived at NLO and NNLO respectively. As we can see from Fig. \ref{profiling1}, the PDFs uncertainties are significantly reduced in the gluon PDF in the full $x$ range and medium $x$ range for the sea quark distribution. 
In addition to the PDF profiling, impact of the measurement on $\alpha_\mathrm{S}$ is investigated at NLO and NNLO obtaining $\alpha_\mathrm{S}(m_\mathrm{Z}) =  0.1170 \pm 0.0018 \,\mathrm{(PDF)} \pm 0.0035 \,\mathrm{(scale)} $ at NLO and $\alpha_\mathrm{S}(m_\mathrm{Z}) =  0.1130 \pm 0.0016 \,\mathrm{ (PDF) } \pm 0.0014 \,\mathrm{ (scale) } $ at NNLO. The NLO result is in good agreement with the world average \cite{ParticleDataGroup:2020ssz}. 

The profiling analysis is repeated by using the triple-differential CMS $\mathrm{t\bar{t}}$ cross section in Ref. \cite{CMS:2019esx} together with the inclusive jets result at NLO. As we can see from the right plot in Fig. \ref{profiling1}, the reduction in the gluon PDF is stronger at high $x$ compared with the result using only inclusive jets on the left plot.
\section{W boson in association with a charm quark}
Associated production of a W boson and a charm (c) quark constitutes a direct probe of the strange (s) PDF of the proton. At the LHC this process has been previously studied at $\sqrt{s}= 7$ and $13$ TeV by the ATLAS \cite{ATLAS:2014jkm} and CMS \cite{CMS:2013wql,CMS:2018dxg} Collaborations, while CMS has recently reported the first results using $8$ TeV data \cite{CMS:2021oxn}. 
The data sample for this study corresponds to an integrated luminosity of $19.7 \,\mathrm{pb}^{-1}$ of proton-proton data collected in 2012. The W boson is identified via its leptonic decay into an isolated muon ($\mu$) or electron ($e$) while c quark jets are tagged in two ways: i) the identification of a muon inside the jet coming from the semileptonic decay of the c quark, and ii) a secondary vertex from a charm hadron decay. 

The $\mathrm{W+c}$ production is dominated by the partonic processes $\mathrm{\bar{s}g} \rightarrow \mathrm{W^{+}}+\mathrm{\bar{c}}$ and $\mathrm{sg}\rightarrow \mathrm{W^{-}}+\mathrm{c}$, then it is characterized by the opposite sign (OS) of the electric charges of the W boson and the c quark. Since for most of the background processes the probabilities of selecting an event with a c quark and a W boson with the same sign (SS) and OS are the same, then the subtraction of OS and SS distributions yields a clean subtraction of backgrounds. 

\begin{figure}
\begin{center}
\includegraphics[width=\textwidth]{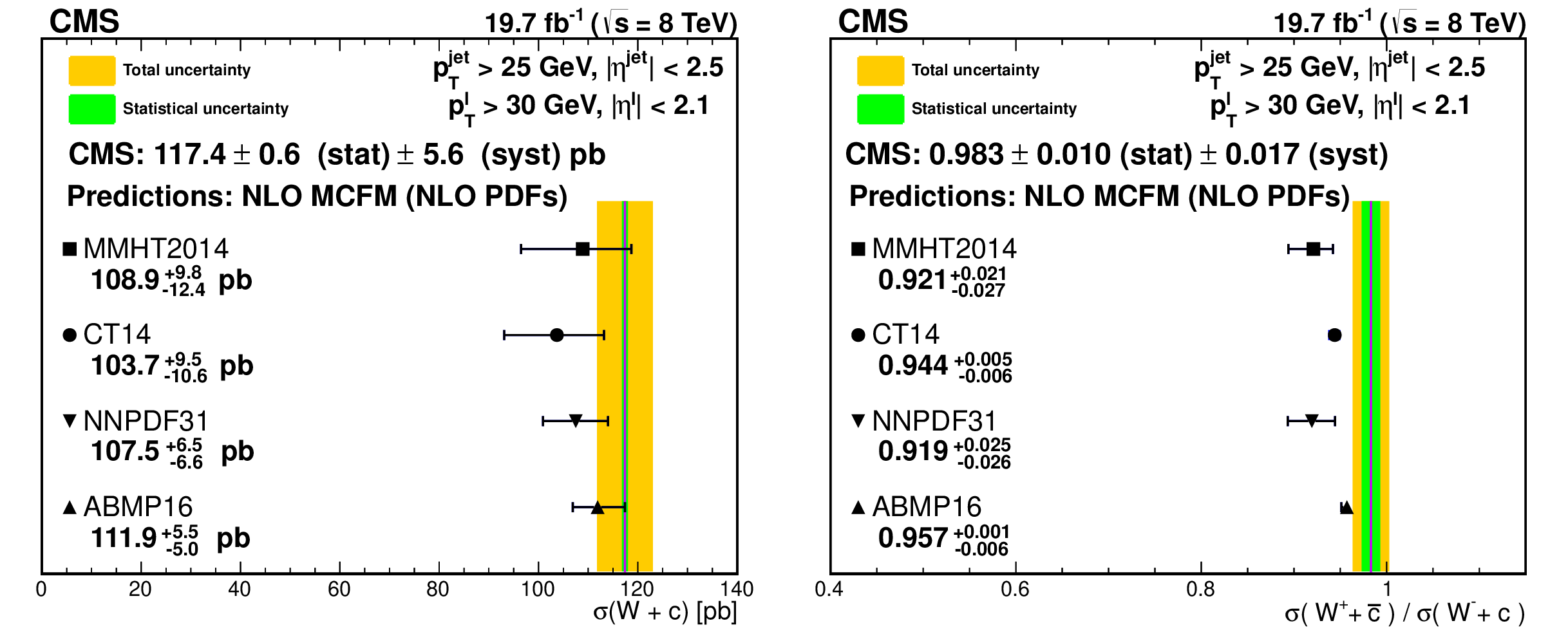}
\caption{Theoretical predictions compared with the measured cross section (left) and cross-section ratio (right). Plots taken from Ref. \cite{CMS:2021oxn}.}
\label{xSec2}
\end{center}
\end{figure}

\begin{figure}
\begin{center}
\includegraphics[width=\textwidth]{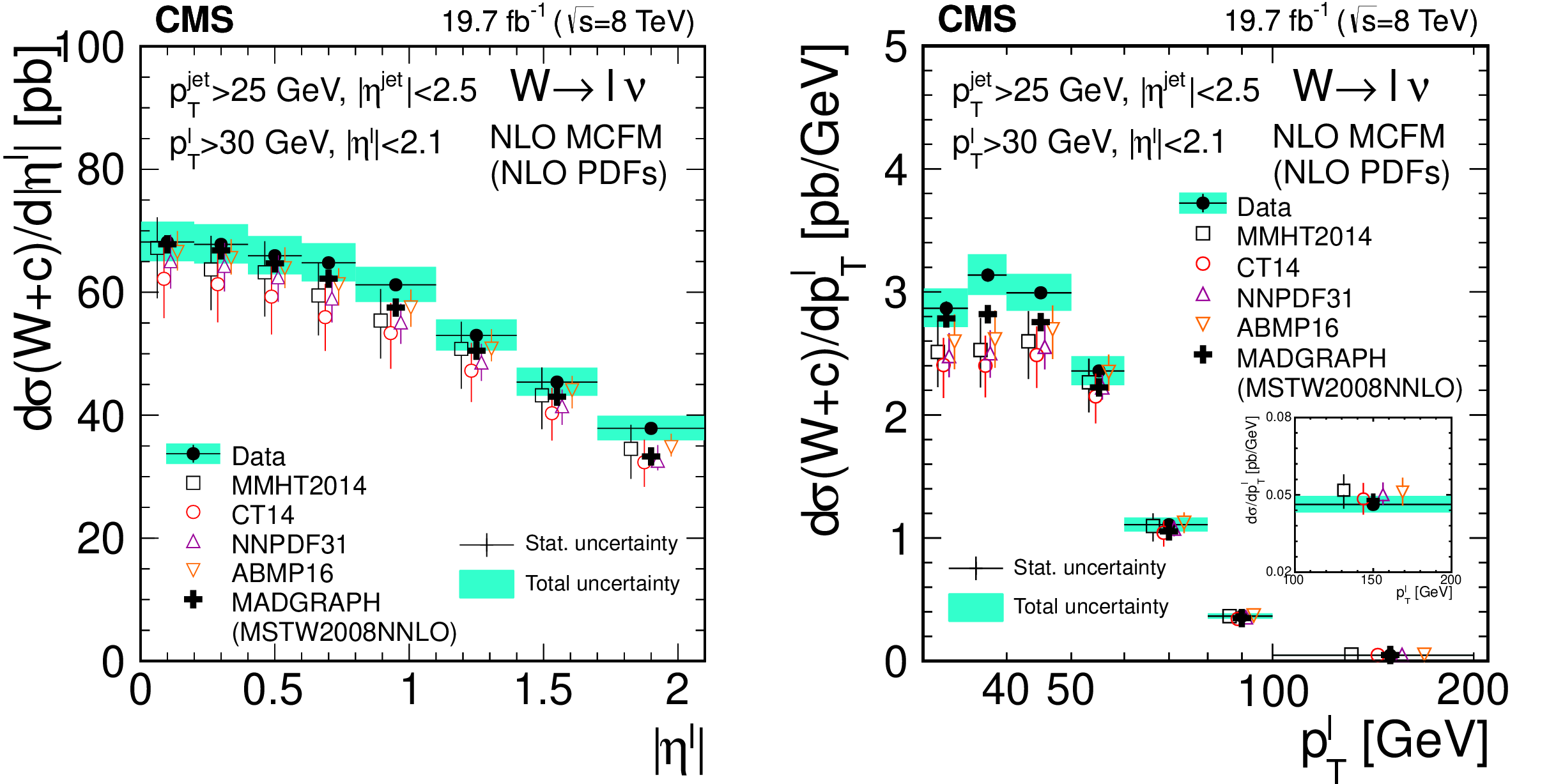}
\caption{Differential cross sections, $\mathrm{d} \sigma(\mathrm{W+c})/\mathrm{d} |\eta^{\mathrm{\ell}}|$ (left) and $\mathrm{d} \sigma(\mathrm{W+c})/\mathrm{d} p_\mathrm{T}^{\mathrm{\ell}}$, compared with theoretical predictions. Plots taken from Ref. \cite{CMS:2021oxn}.}
\label{test}
\end{center}
\end{figure}

The measured inclusive cross section $\sigma(\mathrm{W+c})$, the cross-section ratio $ \sigma(\mathrm{W^{+}+\bar{c}})/\sigma(\mathrm{W^{-}+c}) $ and theoretical predictions are shown in Fig. \ref{xSec2}. The measured $\sigma(\mathrm{W+c})$ cross section is in agreement with theory while the cross-section ratio  is larger than theoretical predictions but within two or three standard deviations. Moreover, differential cross sections are obtained as functions of the absolute value of pseudorapidity ($\eta^{\ell}$) and, for the first time, the transverse momentum ($p_{\mathrm{T}}^{\ell}$) of the lepton coming from the decay of the W boson, as shown in Fig. \ref{test}.

\begin{figure}
\begin{center}
\includegraphics[width=0.45\textwidth]{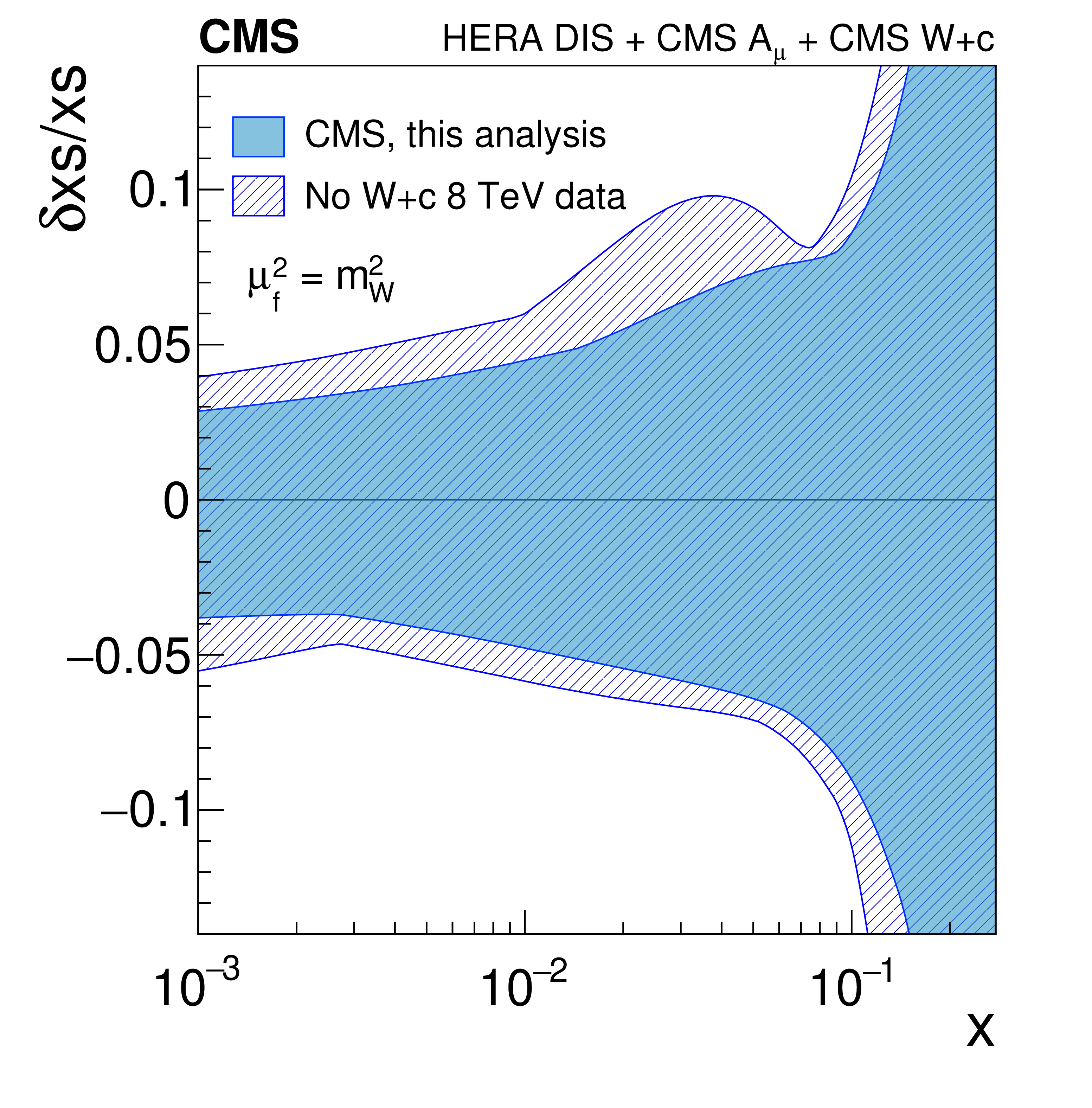}
\includegraphics[width=0.45\textwidth]{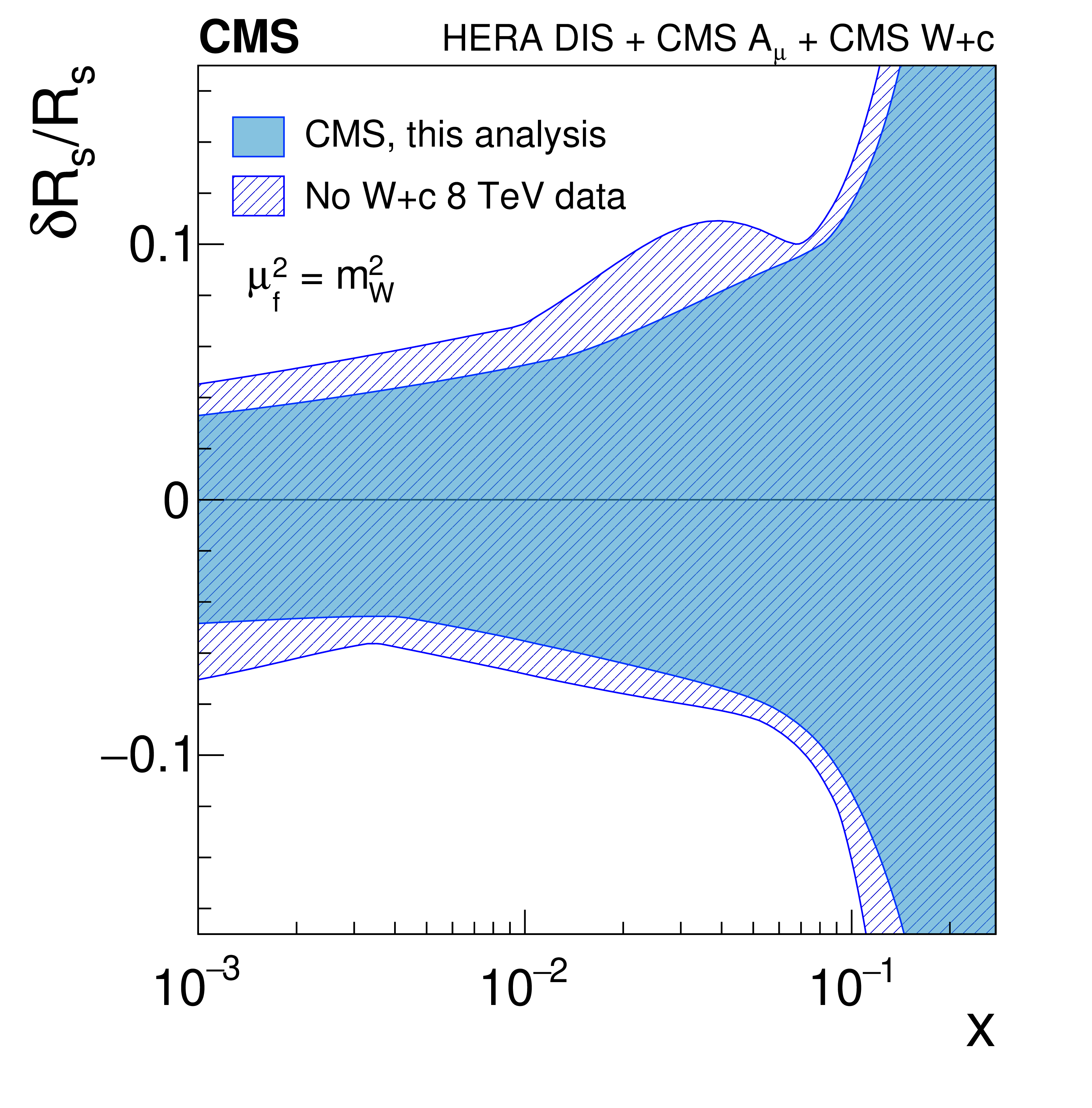}
\caption{Comparison of the relative total uncertainties with and with no $\mathrm{W+c}$ data at $\sqrt{s}=8$ TeV for the strange quark distribution (left) and the strangeness suppression factor (right). Plots taken from Ref. \cite{CMS:2021oxn}.}
\label{impact}
\end{center}
\end{figure}

The present measurement of the $\mathrm{W+c}$ cross section as a function of $|\eta^{\ell}|$ and for lepton $p_{\mathrm{T}}^{\ell}>30$ GeV is used in a QCD analysis at NLO in conjunction with the CMS measurements of $\mathrm{W+c}$ production at $\sqrt{s}=7$ \cite{CMS:2013wql} and 13 TeV \cite{CMS:2018dxg}. Also, the combination of the HERA inclusive deep inelastic scattering cross sections \cite{H1:2015ubc} and the CMS measurements of the lepton charge asymmetry in W boson production at $\sqrt{s}=7$ and 8 TeV \cite{CMS:2013pzl,CMS:2016qqr} are used.  The impact of the $\mathrm{W+c}$ measurement at $\sqrt{s}=8$ TeV on the strange quark distribution $x\mathrm{s}(x,\mu_f^2)$ and the strangeness suppression factor $R_\mathrm{s}(x, \mu_f^2) = (\mathrm{s+\bar{s}})/(\mathrm{\bar{u}+\bar{d}})$ is obtained, as shown in Fig. \ref{impact}, where uncertainty reductions are clearly observed.

\section{Summary}
The CMS Collaboration has presented a measurement of the inclusive jet production cross section in proton-proton collisions at $\sqrt{s}=13$ TeV. The sensitivity of this measurement to the proton PDFs and to the strong coupling constant is assessed in a QCD analysis. The results show a significant reduction of the gluon and sea quark PDFs uncertainties. 

On the other hand, CMS has recently reported results on $\mathrm{W+c}$ production at $\sqrt{s}=8$ TeV and the impact of this measurement on the strange PDF of the proton, showing a significant reduction on the relative uncertainties. 

\bibliographystyle{auto_generated.bst} 
\bibliography{Alcerro.bib}
\end{document}